\documentclass[aps,prx,groupedaddress,reprint,showpacs,superscriptaddress]{revtex4-1}
\usepackage{graphicx}% Include figure files
\usepackage{tabularx}
\usepackage{multirow}
\usepackage{longtable}
\usepackage{dcolumn}% Align table columns on decimal point
\usepackage{bm}% bold math
\usepackage{amsmath}
\usepackage{amssymb}
\usepackage{longtable}
\usepackage{txfonts}
\usepackage{url}
\usepackage[colorlinks=true, linkcolor=blue, anchorcolor=blue, citecolor=blue, urlcolor=magenta]{hyperref}
\usepackage[usenames,dvipsnames]{color}
\definecolor{Gray}{gray}{0.0}
\definecolor{lightGray}{gray}{0.35}

\begin{document}
\title{
%{\color{red}
  High-pressure Mg-Sc-H phase diagram and its superconductivity
  from first-principles calculations 
  %\\
  %or \\
  %First-principles Mg-Sc-H phase diagram under high pressure and its predicted superconductivity
%}
}
\author{Peng Song}
\affiliation{School of Information Science, JAIST, Asahidai 1-1, Nomi, Ishikawa 923-1292, Japan}
\author{Zhufeng Hou}
\affiliation{State Key Laboratory of Structural Chemistry, Fujian Institute of Research on the Structure of Matter, Chinese Academy of Sciences, Fuzhou 350002, China}
\author{Pedro Baptista de Castro}
\affiliation{National Institute for Materials Science, 1-2-1 Sengen, Tsukuba, Ibaraki 305-0047, Japan}
\affiliation{University of Tsukuba, 1-1-1 Tennodai, Tsukuba, Ibaraki 305-8577, Japan}
\author{Kousuke Nakano}
\affiliation{School of Information Science, JAIST, Asahidai 1-1, Nomi, Ishikawa 923-1292, Japan}
\affiliation{International School for Advanced Studies (SISSA), Via Bonomea 265, 34136, Trieste, Italy}
\author{Kenta Hongo}
\affiliation{Research Center for Advanced Computing Infrastructure, JAIST, Asahidai 1-1, Nomi, Ishikawa 923-1292, Japan}
\author{Yoshihiko Takano}
\affiliation{National Institute for Materials Science, 1-2-1 Sengen, Tsukuba, Ibaraki 305-0047, Japan}
\affiliation{University of Tsukuba, 1-1-1 Tennodai, Tsukuba, Ibaraki 305-8577, Japan}
\author{Ryo Maezono}
\affiliation{School of Information Science, JAIST, Asahidai 1-1, Nomi, Ishikawa 923-1292, Japan}

\vspace{10mm}
%\affiliation{$^{*}$
%  mwkumk1702@icloud.com
%}

\date{\today}
%------------------
\begin{abstract}{
%\color{red}
  In this work, global search for crystal structures
  of ternary Mg-Sc-H hydrides (Mg$_x$Sc$_y$H$_z$) under high pressure
  ($100 \le P \le 200$ GPa)
  were performed using the evolutionary algorithm and
  first-principles calculations.
  Based on them, we computed the thermodynamic convex hull and pressure-dependent phase diagram
  of Mg$_x$Sc$_y$H$_z$ for $z/(x+y) < 4$.
  We have identified the stable crystal structures of
  four thermodynamically stable compounds
  with the higher hydrogen content, i.e.,
  $R\bar{3}m$-MgScH$_{6}$, $C2/m$-Mg$_{2}$ScH$_{10}$,
  $Immm$-MgSc$_{2}$H$_{9}$ and $Pm\bar{3}m$-Mg(ScH$_{4}$)$_{3}$.
  Their superconducting transition temperatures were
  computationally predicted by
  the McMillan-Allen-Dynes formula combined
  with first-principles phonon calculations.
  They were found to exhibit superconductivity;
  among them, $R\bar{3}m$-MgScH$_{6}$ was predicted
  to have the highest $T_{c}$ (i.e. 23.34 K) at 200 GPa.}
\end{abstract}
%------------------
\maketitle

%%%%%%%%%%%%%%
\section{Introduction}
\label{sec.intro}
%%%%%%%%%%%%%%%%%%
Hydrogen-rich hydrides of rare earth metals and alkaline earth metals are now recognized to be viable routes to realize room temperature superconductivity under high pressure.~\cite{2017PEN,2016ZUR,2019ZUR,2020SEMa}
These hydrides are usually stabilized with the clathrates consisting of H atoms, which could significantly lower the pressure-volume ($PV$) term of enthalpy and thus preserve the stability at low pressures.~\cite{2012WAN,2015FEN,2014HOO,2021CHE,2018YE,2017QIA,2017PEN,2017LIU,2020KRU}
In these structures, the H atoms can substantially contribute to the electronic density of states around the Fermi level and also the phonon density of states. Such a feature enables them to be the potential candidates to exhibit high-temperature superconductivity. For instance, theoretical calculations recently have predicted that YH$_{10}$ with H$_{32}$ cage structure might exhibit room-temperature superconductivity (286-326 K).~\cite{2017LIU}

%{\color{red}
%  Hereafter, too long paragraphs were splitted into several parts,
%  each of which includes ``one topic'' only.
%}

\vspace{2mm}
%{\color{red}
For binary alkaline earth and rare earth hydrides,
a systematic search for their crystal structures
%}
and superconductivity has been accomplished mostly
by theoretical prediction,
with the exception of few magnetic rare earth hydrides.~\cite{2020SEMa,2020JOS}
%({\color{red}Cite Refs.})
The binary hydrides potentially with high-temperature superconductivity have attracted great attention for the experimental verification. LaH$_{10}$, CaH$_{6}$, YH$_{6}$, BaH$_{12}$ and CeH$_{9}$ are the examples of the superconducting clathrate structures confirmed in experiments.~\cite{2019DRO,2021TRO,2021MA,2021CHE,2019SAL}
The agreement between the theoretical prediction and the experimental discovery in the superconductivity of these binary hydrides has greatly encouraged the theoretical search for metal hydrides in a more extensive range such as the ternary case.

\vspace{2mm}
%{\color{red}
For ternary cases, a very recent experiment on the La-Y-H system
has revealed
%} 
that the synthesized (La,Y)H$_{10}$ at $P = 180$ GPa
exhibits a $T_{c}$ of 253 K.~\cite{2021SEMb} More interestingly,
the pressure required for (La,Y)H$_{10}$ to achieve superconductivity is lower than that of LaH$_{10}$ high-temperature superconductor, indicating that the ternary hydride has more potential in the search for low-pressure room-temperature superconductivity.~\cite{2017LIU,2019DRO}

\vspace{2mm}
The stability analysis and superconductivity prediction for a minor portion of ternary metal hydries, such as La-Y-H, Ca-Y-H, Sc-Ca-H, Sc-Y-H, Y-Mg-H, and Ca-Mg-H, have been accomplished recently by theoretical calculations.~\cite{2021SEMb,2021SONa,2019LIA,2021SHI,2020WEI,2021SONb,2020SUK}
The majority of the high-temperature superconducting compounds in the aforementioned systems prefer to the cage-like structures.
For example, $Fd\bar{3}m$-CaYH$_{12}$ with cubic structure, preserves the clathrate structure consisting of the H$_{24}$ cages, as does CaH$_{6}$ and YH$_{6}$. Meanwhile $Fd\bar{3}m$-CaYH$_{12}$ can remain stable above 170 GPa with a $T_{c}$ value of 254 K as compared to CaH$_{6}$ and YH$_{6}$.~\cite{2019LIA,2012WAN,2017PEN}
Theoretical calculations further revealed that the strong electron-phonon coupling (EPC) in these clathrate structures of ternary metal hydrides is associated strongly to the phonon mode of the H-H bond in the cage.~\cite{2019LIA}
Furthermore, the characteristics of these materials look very similar, including the atomic radius, electron number (spd valence electrons), electron negative, atomic mass of constituent elements and so on.~\cite{2020SEMa} This might prevent the H cage from collapsing and preserve the same cage structure and superconductivity in the ternary hydride with clathrate structure.
ScCaH$_{8}$ and ScCaH$_{12}$ are two potential high-temperature superconductor compounds in the Sc-Ca-H system. The cage structure is preserved in these two compounds and the corresponding $T_{c}$ values are around 212 K and 182 K at 200 GPa, respectively.~\cite{2021SHI}

\vspace{2mm}
Motivated by the aforementioned studies, herein we concentrated our studies on the Mg-Sc-H system to explore the phase diagram and superconductivity of the associated ternary compounds under high pressure. By employing the evolutionary algorithm for crystal structure prediction, we have found the stable structures of MgScH$_{6}$, Mg$_{2}$ScH$_{10}$, MgSc$_{2}$H$_{9}$, and Mg(ScH$_{4}$)$_{3}$ under high pressure, which are expected to have the highest hydrogen content in the ternary Mg-Sc-H compounds. In the studied pressure range (i.e., 100-200 GPa), no stable compounds beyond the hydrogen content in the aforementioned compounds were discovered in the hydrogen-rich cases of Mg-Sc-H system, unlike the La-Y-H, Ca-Y-H, and other ternary systems.~\cite{2021SEMb,2021SONa,2019LIA,2021SHI,2020WEI,2021SONb,2020SUK} Although some of the stable structures of ternary Mg-Sc-H compounds are predicted to exhibit superconductivity, their superconducting transition temperatures are high up to only 23.34 K at 200 GPa, owing to the relatively low density of states at their Fermi level and also a relatively weak electron-phonon coupling (EPC).
%%%%%%%%%%%%%%%%%%
\section{Method}
%%%%%%%%%%%%%%%%%%
%{\color{red}
  We considered the fixed compositions of ternary Mg-Sc-H compounds
  (MgSc$_{3}$H$_{x}$, MgSc$_{2}$H$_{x}$, MgScH$_{x}$, Mg$_{2}$ScH$_{x}$, and Mg$_{3}$ScH$_{x}$,
  where $x$ = 2--12, 14, 16, and 18)
  along selected lines in ternary convex hull at fixed pressure of 100 and 200 GPa. 
  The crystal-structure search for the Mg-Sc-H systems was performed using the evolutionary algorithm implemented in the USPEX (Universal Structure Predictor: Evolutionary Xtallography)~\cite{2006GLA} software, together with the first-principles calculations of structural optimization based on density functional theory (DFT). 
%  }

\vspace{2mm}
%{\color{red}
Our DFT structural optimization 
%}
was carried out using the VASP (Vienna \textit{ab initio} simulation package)~\cite{1993KRE,1994KRE,1996KRE_a,1996KRE_b} code. The electron-ion interaction is described by the projector augmented wave (PAW)~\cite{Blochl1994prb,Kresse1999prb} method. The cutoff energy for plane-wave basis sets was set to 600 eV. The exchange and correlation potential was treated by the Perdew-Burke-Ernzerhof (PBE) functional~\cite{1996PER} within generalized gradient approximation (GGA).

\vspace{2mm}
%{\color{red}
We computed the ternary convex hull of Mg-Sc-H system using the ConvexHull module in scipy, which involves evaluating Gibbs free energies of the predicted crystal structures at a finite temperature.~\cite{2020VIR}
Their computatial details are given in Supplemental Material (SM).
%}
The stable ternary phase was determined by a criteria that its enthalpy of formation should be smaller than the whole convex hull plane, which means such a phase would not decompose into any combination of elementary, binary, or other ternary phases.

\vspace{2mm}
The EPC and phonons of the stable ternary Mg-Sc-H phases were predicted using the QUANTUM ESPRESSO (QE) suite of programs~\cite{2009GIA,2017GIA,2020GIA} with the PAW method and the Perdew-Burke-Ernzerhof (PBE)~\cite{1996PER} exchange correlation function. The cutoff energy for plane-wave basis sets in the QE calculations was set to 80 Ry. The $k$ (and $q$)-point sampling for Broullion zone integration was set as follows: $4\times4\times4$ ($16\times16\times 16$) for MgScH$_{6}$, $5\times 5 \times 2$ ($20\times20\times8$) for Mg$_{2}$ScH$_{10}$, $3\times4\times4$ ($12\times16\times16$) for MgSc$_{2}$H$_{9}$, $5\times5\times5$ ($20\times20\times20$) for Mg(ScH$_{4}$)$_{3}$, respectively. The McMillan-Allen-Dynes formula~\cite{1968MCM} and the Eliashberg function derived from the EPC calculation were used to predict the superconducting critical temperature.

%%%%%%%%%%%%%%%%%%
\section{Results and discussion}
\label{sec.results}
%%%%%%%%%%%%%%%%%%
%{\color{red}
\subsection{First-principles phase diagram of Mg-Sc-H system}
%}
To determine the ternary phase diagram of Mg-Sc-H system under high pressure, we have taken the Mg-H and Sc-H binary systems~\cite{2017QIA,2017PEN,2018YE,2013LON,2015FEN} as well as the simple substances of constituent elements (Mg, Sc, and H)~\cite{2009LIU,2005AKA,2007PIC} as references in the estimation of thermodynamic stability.
It is worth noting that the superconducting structures of the Mg-H and Sc-H binary hydride systems have been discovered recently. For instance, MgH$_{6}$ and ScH$_{7}$ have been predicted to exhibit the highest $T_\mathrm{c}$ values of around 260 K and 169 K in the Mg-H and Sc-H binary system, respectively.~\cite{2017QIA,2017PEN,2018YE,2013LON,2015FEN}
However, the pressure to stabilize these two superconducting compounds are both greater than 300 GPa, which is far beyond the mostly focused high-pressure range (i.e., around 200 GPa) of the widely studied metal hydrides in literature.~\cite{2019LIA,2012WAN,2017PEN,2016EIN}

\begin{figure*}[htbp]
  \begin{center}
    \includegraphics[width=\linewidth]{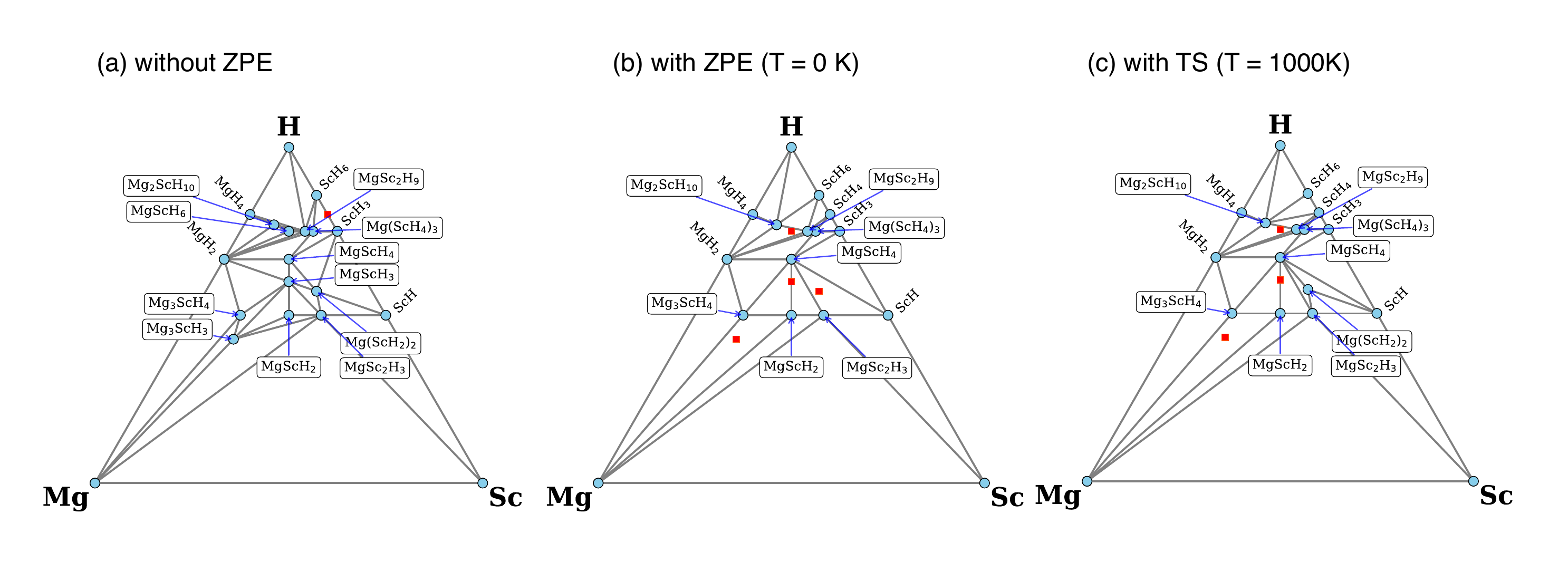}
    \caption{Ternary convex hulls of the Mg-Sc-H system at a pressure of 200 GPa. (b), (c) are the convex hulls of Gibbs free energy at finite temperature. The stable and metastable phases are shown by blue circles and red squares, respectively.}
    \label{fig.convex_hull}
  \end{center}
\end{figure*}

\vspace{2mm}
To explore the most likely composition for the succeeding experimental synthesis, herein we have sampled a wide phase space of the ternary Mg-Sc-H system at fixed pressures (i.e., 100 GPa and 200 GPa).
The ternary phase diagram of the Mg-Sc-H system at 200 GPa is shown in Figure \ref{fig.convex_hull}, while the results for the pressure at 100 GPa and the associated formation energies are given in the SM.
Since most high-pressure studies are carried out under laser heating (1000-2000 K), some metastable phases at 0 K can be stabilized at higher temperatures.~\cite{1986JAY,2021SEMb}	
Therefore, herein we calculate the ternary phase diagram of Mg-Sc-H system at 200 GPa mainly in three different cases: i) formation enthalpy without considering the zero point energy (ZPE), ii) formation enthalpy with the contribution of ZPE and without the contribution of entropy, and iii) Gibbs free energy with the corresponding entropy contribution at 1000 K.

\vspace{2mm}
The stable compounds in the ternary convex hull of Mg-Sc-H system without ZPE at 200 GPa are $Pm\bar{3}m$-Mg$_{3}$ScH$_{3}$, $Pm\bar{3}m$-Mg$_{3}$ScH$_{4}$,  $C2/m$-Mg$_{2}$ScH$_{10}$, $I4_{1}/amd$-MgScH$_{2}$, $P6_{3}/mmc$-MgScH$_{3}$, $P\bar{6}m2$-MgScH$_{4}$, $R\bar{3}m$-MgScH$_{6}$, $Immm$-MgSc$_{2}$H$_{3}$, $R\bar{3}m$-Mg(ScH$_{2}$)$_{2}$, $P6_{3}/mmc$-Mg(ScH$_{2}$)$_{3}$, $Immm$-MgSc$_{2}$H$_{9}$ and $Pm\bar{3}m$-Mg(ScH$_{4}$)$_{3}$.
It is noted that the incorporation of the ZPE contribution leads to the transition of ScH$_{4}$ from a metastable state to a stable one, which is consistent with the results published by Ye et al.~\cite{2018YE} for the binary Sc-H system.

\vspace{2mm}
On the other hand, $Pm\bar{3}m$-Mg$_{3}$ScH$_{4}$, $P6_{3}/mmc$-MgScH$_{3}$, $R\bar{3}m$-MgScH$_{6}$, and $R\bar{3}m$-Mg(ScH$_{2}$)$_{2}$ undergo a change from a stable state to a metastable state because of their large contribution of the ZPE.
The four ternary compounds are now equal to the convex hull plane, with energy differences of 0.003, 0.032, 0.008 and 0.015 eV/atom, respectively.
Furthermore, we noticed that at a high temperature of 1000 K, $R\bar{3}m$-Mg(ScH$_{2}$)$_{2}$ can shift from a metastable state to a stable one due to the entropy contribution of lattice vibration.
In the binary Mg-H and Sc-H systems, $Cmcm$-MgH$_{4}$ and $Cmcm$-ScH$_{6}$ are the stable compounds with the greatest H content in the pressure range of 100-200 GPa.~\cite{2018YE,2015FEN}
Despite the fact that most of the existing high-temperature superconducting hydrides are hydrogen-rich compounds, no stable compounds in the extremely hydrogen-rich case have been identified in the pressure range of 100--200 GPa we searched.
The ternary diagram package contains the synthetic routes for these stable compounds, and the connections of the binary and unary systems can be used as prospective candidates for the experimental synthesis of the ternary system.

\begin{figure}[htbp]
  \begin{center}
    \includegraphics[width=0.8\linewidth]{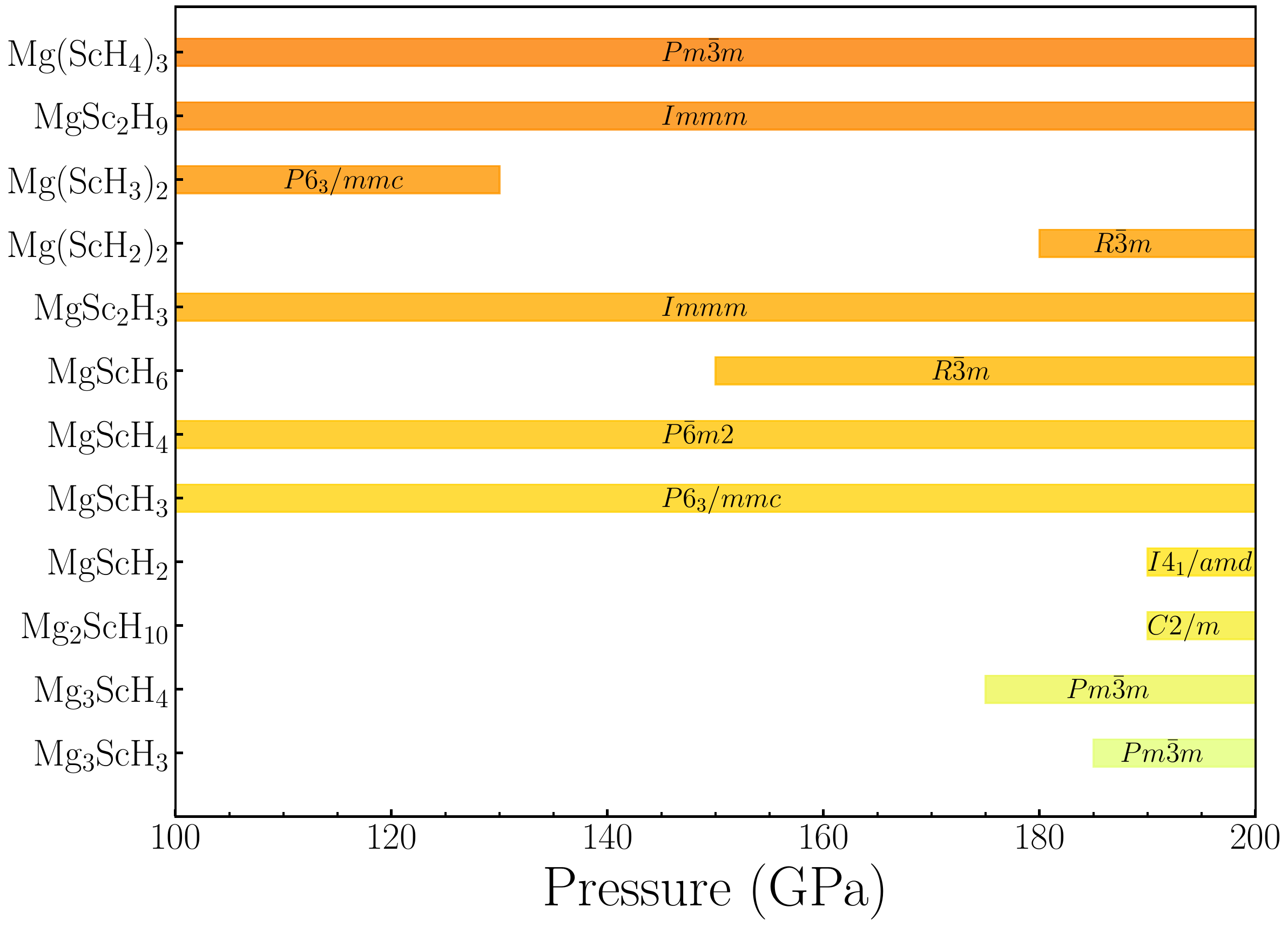}
    \caption{Phase diagram of Mg-Sc-H in the pressure range of 100--200 GPa.}
    \label{fig.phase_diagram}
  \end{center}
\end{figure}

\vspace{2mm}
Based on the calculated formation of enthalpy without accounting for the ZPE contribution, we summarize the pressure-dependent phase stability of Mg-Sc-H system and present it in Figure~\ref{fig.phase_diagram}. In particular, several phases including $Pm\bar{3}m$-Mg(ScH$_{4}$)$_{3}$, $Immm$-MgSc$_{2}$H$_{10}$, $P\bar{6}m2$-MgScH$_{4}$, and $P6_{3}/mmc$-MgScH$_{3}$ can be stable in a wide pressure range from 100 GPa to 200 GPa. It is remarkably to note that the Mg-rich ternary Mg-Sc-H phases prefer to be stable in the pressure range toward 200 GPa, while the Sc-rich ones except $R\bar{3}m$-Mg(ScH$_2$)$_2$ can also be stabilized at the pressure of 100 GPa. $P6_{3}/mmc$-Mg(ScH$_2$)$_2$ can be stable only in the lower pressure range of 100--130 GPa.
This trend might be associated with to the pressure range of stabilized binary Mg-H and Sc-H hydride systems. At the low pressure, the number of possible stable compounds in the Sc-H system are more than that of the Mg-H system.~\cite{2018YE,2015FEN}

%{\color{red}
  \subsection{Predicted superconductivity of Mg-Sc-H system}
%  }
\vspace{2mm}
Previous studies on the alkali metals and alkaline earth hydrides have shown that the superconducting properties of these hydrides are strongly dependent on the hydrogen content and that the higher $T_\mathrm{c}$ is more likely to appear in the hydrogen-rich cases.~\cite{2020SEMa,2017PEN} Herein we pay more attention to $R\bar{3}m$-MgScH$_{6}$, $C2/m$-Mg$_{2}$ScH$_{10}$, $Immm$-MgSc$_{2}$H$_{9}$, and $Pm\bar{3}m$-Mg(ScH$_{4}$)$_{3}$) for their electronic structures and superconductivity because they contain very high hydrogen content.

\begin{figure}[htbp]
  \begin{center}
    \includegraphics[width=\linewidth]{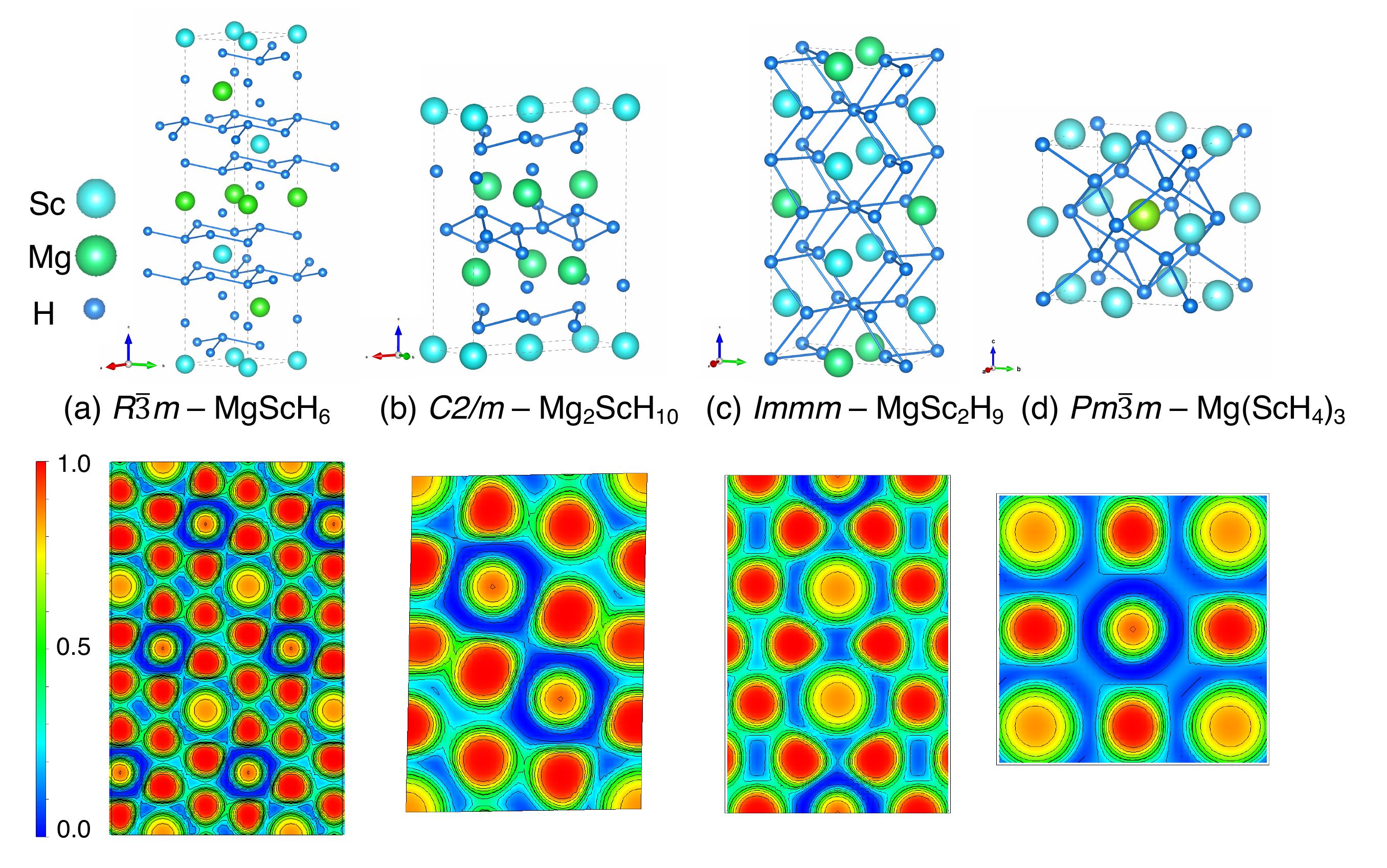}
    \caption{Predicted crystal structures of (a) $R\bar{3}m$-MgScH$_{6}$,  (b) $C2/m$-Mg$_{2}$ScH$_{10}$,  (c) $Immm$-MgSc$_{2}$H$_{9}$,  (d) $Pm\bar{3}m$-MgSc$_{3}$H$_{12}$ at 200 GPa. Bottom panels show the corresponding contour plots of electron localization function (ELF) in these structures. These structural models were drawn by using VESTA.~\cite{2011MOM}}
    \label{fig.crystal_structure}
  \end{center}
\end{figure}

\vspace{2mm}
The crystal structures and the electron localization function (ELF) of these aforementioned four compounds are shown in Figure~\ref{fig.crystal_structure}.
$R\bar{3}m$-MgScH$_{6}$ is stable above 150 GPa and has a hexagonal crystal structure, in which each Mg and Sc atom is surrounded by 14 H atoms.
The shortest H-H bond length in this structure is about 1.587 $\AA$ and the corresponding ELF value at the H-H bond center is about 0.4, indicating a nearly metallic character.
The ultra-short distance between atoms translates to a significant ELF value of around 0.7, indicating that the H-H bond in $C2/m$-Mg$_{2}$ScH$_{10}$ exhibits a strong covalent character.
In $Immm$-MgSc$_{2}$H$_{9}$ and $Pm\bar{3}m$-Mg(ScH$_{4}$)$_{3}$, the shortest H-H bond lengths are 1.628 $\AA$ and 1.650 $\AA$, respectively, and the corresponding ELF values at the H-H bond center are 0.3 and 0.1. However the H lattice sites exhibit much greater ELF values. Therefore, the chemical bonding in $Immm$-MgSc$_{2}$H$_{9}$ and $Pm\bar{3}m$-Mg(ScH$_{4}$)$_{3}$ is mostly ionic interaction.
Furthermore, $Pm\bar{3}m$-Mg(ScH$_{4}$)$_{3}$ can still form a clathrate structure, in which each Mg and Sc atom is surrounded by the H14 cages. This structure may be regarded as a structure created when the central Sc in $Fm\bar{3}m$-ScH$_{3}$ is replaced by Mg.

\vspace{2mm}
We should point out that $Fm\bar{3}m$-ScH$_{3}$ does not exhibit superconductivity, as reported previously by Ye et al,~\cite{2018YE}. Therefore, the partial substitution of Sc in $Fm\bar{3}m$-ScH$_{3}$ by Mg triggers the superconductivity in $Pm\bar{3}m$-Mg(ScH$_{4}$)$_{3}$.

\begin{figure}[htbp]
  \begin{center}
    \includegraphics[width=\linewidth]{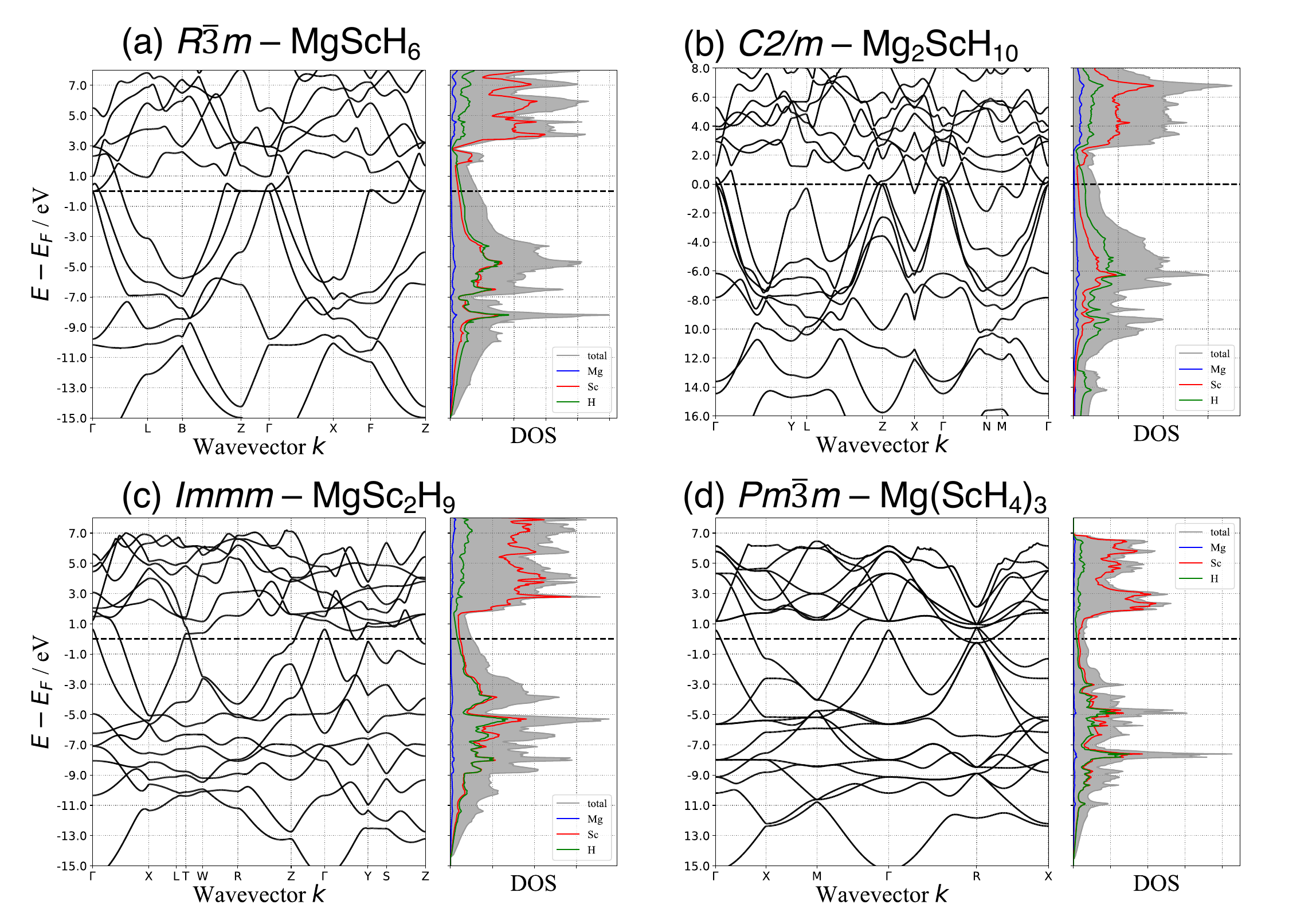}
    \caption{Electronic band structures and atom-projected electronic density of states of (a) $R\bar{3}m$-MgScH$_{6}$,  (b) $C2/m$-Mg$_{2}$ScH$_{10}$,  (c) $Immm$-MgSc$_{2}$H$_{9}$,  (d) $Pm\bar{3}m$-MgSc$_{3}$H$_{12}$ at 200 GPa.}
    \label{fig.band_structure}
  \end{center}
\end{figure}

\vspace{2mm}
Figure~\ref{fig.band_structure} shows the calculated electronic energy band structure and atom-projected electronic density of states (eDOS) of the aforementioned four structures of great interest. All of them at 200 GPa exhibit metallic feature. Among these four structures, $R\bar{3}m$-MgScH$_{6}$ has most robust density of states at the Fermi level ($E_\mathrm{F}$), while $Pm\bar{3}m$-MgSc$_3$H$_{12}$ has least density of states at $E_\mathrm{F}$. The dominated contributions to the density of states at $E_\mathrm{F}$ come from the Sc and H atoms, while the contribution from Mg atoms is negligible. The valence bands in all of these four structures are mainly ascribed to the strong hybridization between Sc and H atoms. At the $E_\mathrm{F}$ of $R\bar{3}m$-MgScH$_{6}$, there are doubly degenerated bands with very flat feature along the $\Gamma \to$ $Z$ direction (i.e., parallel to the $c$ axis of the hexagonal unit cell of $R\bar{3}m$-MgScH$_{6}$). Such a feature of flat band would aid in the electron-phonon interaction.~\cite{1997SIM} This also lead to the highest $T_\mathrm{c}$ of $R\bar{3}m$-MgScH$_{6}$ among these four structures.
	
\begin{figure}[htbp]
  \begin{center}
    \includegraphics[width=\linewidth]{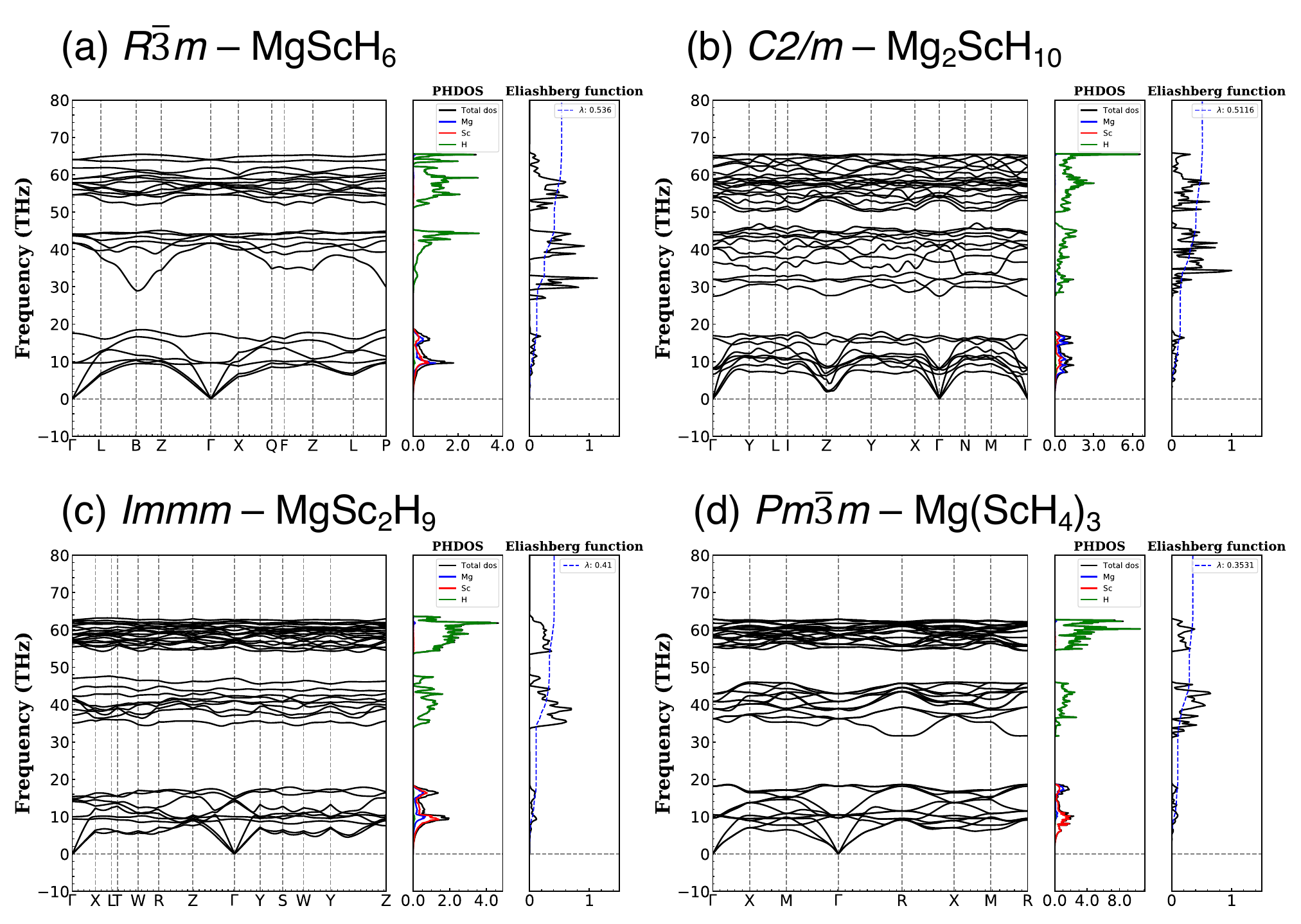}
    \caption{Phonon dispersion and atom-projected phonon density of states (pDOS), and Eliashberg spectral  of (a) $R\bar{3}m$-MgScH$_{6}$,  (b) $C2/m$-Mg$_{2}$ScH$_{10}$,  (c) $Immm$-MgSc$_{2}$H$_{9}$,  (d) $Pm\bar{3}m$-MgSc$_{3}$H$_{12}$ at 200 GPa. }
    \label{fig.phonon_structure}
  \end{center}
\end{figure}

\vspace{2mm}
Figure \ref{fig.phonon_structure} shows the phonon band structure, atom-decomposed phonon density of states (pDOS), and Eliashberg spectra of $R\bar{3}m$-MgScH$_{6}$, $C2/m$-Mg$_{2}$ScH$_{10}$, $Immm$-MgSc$_{2}$H$_{9}$, and $Pm\bar{3}m$-Mg(ScH$_{4}$)$_{3}$ at 200 GPa.
At first, we can see that there are no imaginary phonons in all of these four structures and so they are dynamically stable. In addition, their phonons can be clearly grouped into three frequency regions. The low frequency region (i.e., below 20 THz) is dominantly ascribed to the vibration of Mg and Sc atoms because of their heavier atomic masses. The middle frequency region (i.e., centered around 40 THz) and the high frequency region (i.e., above 50 THz) are exclusively contributed by the vibration of H atom.

\vspace{2mm}
%{\color{red}
Table~\ref{table.Pdep} lists $T_c$ values and their related quantities.
%}
The estimated EPC constants ($\lambda$) of these four structures follow the ordering of $R\bar{3}m$-MgScH$_{6}$ $>$ $C2/m$-Mg$_{2}$ScH$_{10}$ $>$ $Immm$-MgSc$_{2}$H$_{9}$ $>$ $Pm\bar{3}m$-Mg(ScH$_{4}$)$_{3}$. Their $T_\mathrm{c}$ values exhibit the same ordering. $R\bar{3}m$-MgScH$_{6}$ is found to have the highest $T_\mathrm{c}$ (i.e., 23.3 K) among these four structures at 200 GPa. However, this superconducting temperature is much lower than the reported other metal hydrides with high hydrogen content such as LaH$_{10}$ and YH$_{10}$.~\cite{2020SEMa,2017PEN}
This is due to the fact that the contribution of hydrogen atoms to the electronic density of states at $E_\mathrm{F}$ of these four structure is too low and the overall EPC constant is also too small. The contribution of heavy atoms (Mg and Sc) to the overall EPC constant in the low frequency range is extremely small, accounting for only about 8\%.

\vspace{2mm}
It is interesting to note that YH$_{3}$ and $Fm\bar{3}m$-ScH$_{3}$ have the same hydrogen content per metal atom with $R\bar{3}m$-MgScH$_{6}$, $Immm$-MgSc$_{2}$H$_{9}$, and $Pm\bar{3}m$-Mg(ScH$_{4}$)$_{3}$.  However, YH$_{3}$ has a strong EPC constant of 1.6 and a $T_\mathrm{c}$ value of around 40 K.~\cite{2009KIM} In contrast, the EPC constant of $Fm\bar{3}m$-ScH$_{3}$ is about 0.23 and it hardly leads to the superconductivity.~\cite{2018YE} The EPC constants of aforementioned four structures of Mg-Sc-H are in the range from 0.35 to 0.54, and thus their superconducting transition temperatures are lower than that of YH$_3$.

%%%%%%%%%%%%%%%%%%%%%%%%%
\section{Conclusion}
\label{sec.conc}
%------------------
In summary, we have employed the evolutionary algorithm in the USPEX code and the first-principles calculations to explore the ternary phase diagram of the Mg-Sc-H system under pressure from 100 to 200 GPa.
%The stable pressure range shows a certain relationship with the Mg/Sc ratio.
Our calculations show that the Sc-rich ternary Mg-Sc-H compounds are favorable to be stable around 100 GPa, while the Mg-rich ones are more likely to stable above 180 GPa.
%This is because the Sc-rich ternary compound have lower free energies. The exceptionally strong H-H covalent bond, which could lower the contribution of PV term to the enthalpy, is the primary explanation for Mg-rich $C2/m$-Mg$_{2}$ScH$_{10}$'s stability.
The ternary Mg-Sc-H hydrides with the hydrogen/metal ratio around three have been predicted to exhibit the superconductivity. In particular, the superconducting transition temperature of $R\bar{3}m$-MgScH$_{6}$ is around 23 K at 200 GPa. Although the superconducting transition temperature of $Pm\bar{3}m$-Mg(ScH$_{4}$)$_{3}$ is relatively low, it can be stable in the low range of high pressure toward 100 GPa. Our results provide useful information for the discovery of new low-pressure superconducting hydrides.

%%%%%%%%%%%%%%%%%%
\section*{Acknowledgments}
%%%%%%%%%%%%%%%%%%
The computations in this work have been performed
using the facilities of
Research Center for Advanced Computing
Infrastructure (RCACI) at JAIST.
K.H. is grateful for financial support from 
the HPCI System Research Project (Project ID: hp190169) and 
MEXT-KAKENHI (JP16H06439, JP17K17762, JP19K05029, and JP19H05169)
and the Air Force Office of Scientific Research
(Award Numbers: FA2386-20-1-4036).
R.M. is grateful for financial supports from
MEXT-KAKENHI (19H04692 and 16KK0097),
FLAGSHIP2020 (project nos. hp1
90169 and hp190167 at K-computer),
Toyota Motor Corporation, I-O DATA Foundation,
the Air Force Office of Scientific Research
(AFOSR-AOARD/FA2386-17-1-4049;FA2386-19-1-4015),
and JSPS Bilateral Joint Projects (with India DST).

\bibliographystyle{apsrev4-1}
\bibliography{references}

%%%%%%%%%%%%%%%%%%
%%%%%%%%%%%%%%%%%%
%%%%%%%%%%%%%%%%%%
%------------------
%------------------
%------------------
%------------------

%------------------
%------------------

%------------------
%------------------

%------------------
%------------------

%------------------
\begin{table*}[htbp]
 \begin{center}
   \caption{
     $T_c$ estimated by McMillan formula using
     first-principles phonon calculations
     for
 	Mg-Sc-H
     at each pressure.
     $\lambda$ and $\omega_{\rm log}$ are
     the parameters appearing in the formula.
   }
     \label{table.Pdep}
\begin{tabular}{lcccccc}
\hline
\hline
 %  \specialrule{0em}{1pt}{1pt}
Phase &Space group & $P$ & $\lambda$ & $\omega_{\rm log}$ & $N_{E_\mathrm{F}}$  &$T_\mathrm{c}$~(K)  \\
 &   &~(GPa)           &   & ~(K)   & (states/eV/\AA$^{3}$)&at $\mu$ = 0.1 -- 0.13\\
 \hline
MgScH$_{6}$ & $R\bar{3}/m$  & 200        & 0.536    & 1478.84& 0.0150 &  23.34 -- 15.08 \\
% \hline
Mg$_{2}$ScH$_{10}$ & $C2/m$  & 200        & 0.512   & 1350.17 & 0.0125 &  17.94 --   11.10 \\
%\hline
MgSc$_{2}$H$_{9}$ & $Immm$  & 200        & 0.410   & 1427.39 & 0.0120 &  6.88 --  3.14  \\
%\hline
MgSc$_{3}$H$_{12}$ & $Pm\bar{3}m$ & 200     & 0.353 & 1377.73 &  0.0102 & 2.61 -- 0.83 \\
 \hline
 \hline
\end{tabular}
 \end{center}
\end{table*}
%------------------

%------------------
%\begin{figure}[htbp]
%  \begin{center}
%    \includegraphics[width=0.8\linewidth]{./img/$T_\mathrm{c}$_p}
%    \caption{The predicted crystal structure of (a) $R\bar{3}m$-MgScH$_{6}$,  (b) $C2/m$-Mg$_{2}$ScH$_{10}$,  (c) $Immm$-MgSc$_{2}$H$_{9}$,  (d) $Pm\bar{3}m$-MgSc$_{3}$H$_{12}$.}
%    \label{fig.crystal_structure}
%  \end{center}
%\end{figure}
%------------------

%------------------
%%%%%%%%%%%%%%%%%%%%%%%%%%%%%%%
\end{document}

\title{{\Large Supplemental Material}\\
  \vspace{3mm}
  for\\
  \vspace{3mm}
First-principles study on the phase diagram and superconductivity of Mg-Sc-H system under high pressure}

\author{Peng Song}
\affiliation{School of Information Science, JAIST, Asahidai 1-1, Nomi, Ishikawa 923-1292, Japan}
\author{Zhufeng Hou}
\affiliation{State Key Laboratory of Structural Chemistry, Fujian Institute of Research on the Structure of Matter, Chinese Academy of Sciences, Fuzhou 350002, China}
\author{Pedro Baptista de Castro}
\affiliation{National Institute for Materials Science, 1-2-1 Sengen, Tsukuba, Ibaraki 305-0047, Japan}
\affiliation{University of Tsukuba, 1-1-1 Tennodai, Tsukuba, Ibaraki 305-8577, Japan}
\author{Kousuke Nakano}
\affiliation{School of Information Science, JAIST, Asahidai 1-1, Nomi, Ishikawa 923-1292, Japan}
\affiliation{International School for Advanced Studies (SISSA), Via Bonomea 265, 34136, Trieste, Italy}
\author{Kenta Hongo}
\affiliation{Research Center for Advanced Computing Infrastructure, JAIST, Asahidai 1-1, Nomi, Ishikawa 923-1292, Japan}
\author{Yoshihiko Takano}
\affiliation{National Institute for Materials Science, 1-2-1 Sengen, Tsukuba, Ibaraki 305-0047, Japan}
\affiliation{University of Tsukuba, 1-1-1 Tennodai, Tsukuba, Ibaraki 305-8577, Japan}
\author{Ryo Maezono}
\affiliation{School of Information Science, JAIST, Asahidai 1-1, Nomi, Ishikawa 923-1292, Japan}

%\email{spacesongpy@gmail.com}

\maketitle
%%%%%%%%%%%%%%%%%%%%%%%%%%%%%%%%%%%%%%%%%%%%%%%%%%%%%%%%%%%%%
%%%%%%%%%%%%         experimental and DFT          %%%%%%%%%%
%%%%%%%%%%%%%%%%%%%%%%%%%%%%%%%%%%%%%%%%%%%%%%%%%%%%%%%%%%%%%
%\section{Structural information of the most stable and lowest-lying metastable structures in \ce{W2C}}

\subsection{Computational details}
\label{computational}
%%%%%%%%%%%%%%%%%%
The Gibbs free energy of the chemical compounds in the Sc-Mg-H system was calculated according to the following equation:
\begin{equation}
G(T) = F(T) + pV,
\end{equation}
where $F(T)$ is the Helmholtz free energy as defined below:
\begin{equation}
F(T) = U(T) - TS(T).
\end{equation}
The internal energy $U(T)$ and entropy $S(T)$ are given by the following equations:
\begin{equation}
U(T) = E_\mathrm{elec} + \int_{0}^{\infty } \left[\frac{\varepsilon}{e^{\frac{\varepsilon}{\kappa_{B} T } }  - 1}  +  \frac{\varepsilon}{2}\right]\sigma(\varepsilon)d\varepsilon,
\end{equation}

\begin{equation}
S(T) = \int_{0}^{\infty}\left[\frac{\varepsilon}{T}\frac{1}{e^{\frac{\varepsilon}{\kappa_{B}T}} - 1}  - \kappa_{B} \rm{In}(1 - e^{-\frac{\varepsilon}{\kappa_{B}T}})\right]\sigma(\varepsilon)d\varepsilon.
\end{equation}
Here $\sigma(\varepsilon)$ represents the phonon density of states as a function of vibrational energy and is obtained by the lattice dynamics calculation using the Phonopy code.~\cite{2015TOG} $E_\mathrm{elec}$ is the electronic energy from the DFT calculations.

The superconducting critical temperature was predicted according to the McMillan-Allen-Dynes formula:~\cite{1968MCM}
\begin{equation}
T_\mathrm{c} = \frac{\omega_{log}}{1.2} \rm{exp}\left(\frac{-1.04 (1 + \lambda)}{\lambda (1 - 0.62\mu^{\ast}) - \mu^{\ast}}\right),
\end{equation}
where $\omega_{log}$ and $\lambda$ are the logarithmic average phonon frequency and electron-phonon coupling (EPC) constant obtained in the Eliashberg function.
\clearpage
\newpage
%%%%%%%%%%%%%%%%%%
\subsection{ Supplementary figures}
\label{supp_figure}
%%%%%%%%%%%%%%%%%%
%------------------
%------------------
\begin{figure*}[htbp]
  \begin{center}
    \includegraphics[width=\linewidth]{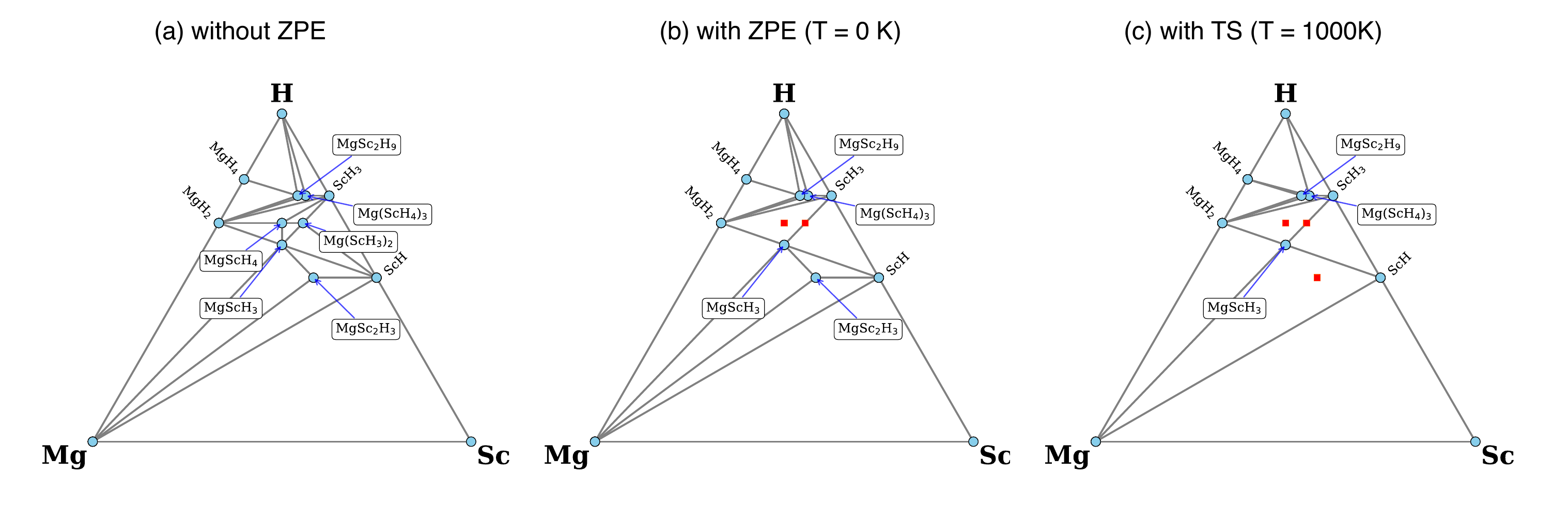}
    \caption{Ternary convex hulls of the Mg-Sc-H system at a pressure of 100 GPa.}
    \label{fig.convex_hull}
  \end{center}
\end{figure*}
%------------------
%------------------
\clearpage
\newpage
\subsection{ Supplementary tables}

\begin{table}[htbp]
\caption{The Gibbs free energy and enthalpy value of Mg-Sc-H at 100GPa at a finite temperature.The metastable phase is shown by the red text in the figure, and the energy value above the convex hull is indicated by the number in brackets.}
\begin{tabular}{cccccccc}
\hline
\hline
Formula   & Space group & Nites & PV  & Enthalpy  & ZPE  & G (T = 0 K) & G (T = 1000 K) \\
   &  &  & (eV/atom) &  (eV/atom) &  (eV/atom) & (eV/atom) & (eV/atom) \\
\hline
Mg        & $Im\bar{3}m$ & 1     & 14.0401      & 7.57398          & 0.07276       & 7.64674    & 7.4084       \\
Sc        & $P6_{2}22$       & 3     &  14.964     &3.82559       & 0.03295        & 3.85854    & 3.55277       \\
H         & $C2/c$       & 12    & 2.8488       & -1.20802            & 0.27219      & -0.93583   & -1.03683        \\
MgH2      & $P6_{3}/mmc$ & 6     & 5.677      & 0.83324           & 0.21098   & 1.04422     &0.94367      \\
MgH4      & $Cmcm$       & 10    & 4.4777       & -0.02043            & 0.21336       &0.19293     & 0.10469        \\
ScH       & $Fm\bar{3}m$ & 2     & 8.7637      & 0.17971            & 0.16433       & 0.34405    & 0.14192        \\
ScH3      & $Fm\bar{3}m$ & 4     & 5.5659      & -1.24658        & 0.20921  & -1.03737& -1.13512    \\
MgScH3    & $P6_{3}/mmc$ & 10    &  7.0434        & 0.39592           & 0.18754       & 0.58346   & 0.46915       \\
MgScH4    & $P\bar{6}m2$ & 6     & 6.1657       & -0.07849            & 0.25699      & {\color{red}0.1785 (+0.02)}     &{\color{red}0.06299 (+0.01)}      \\
MgSc2H3   & $Immm$       & 6     &  8.3424     & 0.87244           & 0.16994        & 1.04238 &{\color{red}0.89664(+0.013)}    \\
Mg(ScH3)2 & $P6_{3}/mmc$  & 18    & 6.3831      & -0.33496         & 0.19989      &{\color{red}-0.13507 (+0.002)}    & {\color{red}-0.24167 (+0.002)}        \\
MgSc2H9   & $Immm$       & 12    &  5.3076       & -0.77339           & 0.2173   & -0.5561    & -0.64476      \\
Mg(ScH4)3 & $Pm\bar{3}m$ & 16    &  5.3679       & -0.90105s            & 0.21949      & -0.68156    & -0.77673       \\
\hline
\hline
\end{tabular}

\end{table}

\begin{table}[htbp]
\caption{200GPa}
\begin{tabular}{cccccccc}
\hline
\hline
Formula   & Space group & Nites & PV  & Enthalpy  & ZPE  & G (T = 0 K) & G (T = 1000 K) \\
   &  &  & (eV/atom) &  (eV/atom) &  (eV/atom) & (eV/atom) & (eV/atom) \\
\hline
Mg        & $Im\bar{3}m$ & 1     & 11.225       & 13.793            & 0.088       & 13.881    & 13.678       \\
Sc        & $C222$       & 3     & 11.385      & 10.272           & 0.042        & 10.314    & 10.062       \\
H         & $C2/c$       & 12    & 2.15       & 0.0182            & 0.293       & 0.311     & 0.2417        \\
MgH2      & $P6_{3}/mmc$ & 6     & 4.598       & 3.372             & 0.267       & 3.639     & 3.562        \\
MgH4      & $Cmcm$       & 10    & 3.578       & 1.964            & 0.276       & 2.24     & 2.173        \\
ScH       & $Fm\bar{3}m$ & 2     & 7.124      & 4.103            & 0.209      & 4.312     & 4.169        \\
ScH3      & $Fm\bar{3}m$ & 4     & 4.654       & 1.282            & 0.258       & 1.540      & 1.457        \\
ScH4      & $I4/mmm$     & 5     & 3.956       &  {\color{red}1.022 (+0.007) }             & 0.233       & 1.255     & 1.176        \\
ScH6      & $Cmcm$       & 14    & 3.462       & 0.711            & 0.263       & 0.974      & 0.907        \\
Mg3ScH3   & $Pm\bar{3}m$ & 7     & 7.149       & 6.464             & 0.205       & 6.669     & {\color{red}6.543 (+0.02) }        \\
Mg3ScH4   & $Pm\bar{3}m$ & 7     & 6.384      & 5.415             & 0.196       & {\color{red} 5.611 (+0.015)}     & 5.491         \\
Mg2ScH10  & $C2/m$       & 13    & 3.999       & 1.888           & 0.261      & 2.149     & 2.074        \\
MgScH2    & $I4_{1}/amd$ & 8     & 6.598       & 4.922            & 0.205       & 5.127     & 5.009        \\
MgScH3    & $P6_{3}/mmc$ & 10    & 5.722        & 3.551            & 0.289        & {\color{red}3.84 (+0.032)}     & {\color{red}3.743 (+0.036)}        \\
MgScH4    & $P\bar{6}m2$ & 6     & 5.069      & 2.7             & 0.229       & 2.928     & 3.743        \\
MgScH6    & $R\bar{3}m$  & 8     & 4.295       & 1.843            & 0.228       & {\color{red}2.118  (+0.008) }       & {\color{red}2.043 (+0.009)}        \\
MgSc2H3   & $Immm$       & 6     & 6.764       & 4.602           & 0.22        & 4.822     & 4.699        \\
Mg(ScH2)2 & $R\bar{3}m$  & 14    & 6.101       & 3.702            & 0.238       & {\color{red} 3.94 (+0.003)}     & 3.823        \\
MgSc2H9   & $Immm$       & 12    & 4.406       & 1.629            & 0.268       & 1.897     & 1.821        \\
Mg(ScH4)3 & $Pm\bar{3}m$ & 16    & 4.464       & 1.532            & 0.266       & 1.799     & 1.722       \\
\hline
\hline
\end{tabular}

\end{table}

%%%%%%%%%%%%%%%%%%

\begin{center}
\begin{longtable}[htbp]{c c c c lccc}
\caption[aaaa]{Crystal structures of Mg-Sc-H predicted at each pressure~($P$). Lattice parameters ($a$, $b$ and $c$) are given in unit of $\AA$.}\\ \hline
\endfirsthead
\multicolumn{8}{c}%
{{\bfseries \tablename\ \thetable{} -- continued from previous page}} \\
\hline \multicolumn{1}{c}{Compound} &
\multicolumn{1}{c}{Space group} &
\multicolumn{1}{c}{$P$~(GPa)}&
\multicolumn{1}{c}{Lattice parameters} &
\multicolumn{4}{c}{Atomic coordinates (fractional)}\\
&&&&   \multicolumn{1}{c}{Atoms} &\multicolumn{1}{c}{$x$}& \multicolumn{1}{c}{$y$} &\multicolumn{1}{c}{$z$}\\
\hline
\endhead
\hline \multicolumn{8}{r}{{Continued on next page}} \\ \hline
\endfoot
\hline \hline
\endlastfoot
Compound & Space group & $P$~(GPa)& Lattice parameters & \multicolumn{4}{c}{Atomic coordinates (fractional)}\\
&&&&    Atoms  & $x$ &  $y$  & $z$ \\
\hline
\specialrule{0em}{1pt}{1pt}
Mg$_{3}$ScH$_{3}$    & $Pm\bar{3}m$ & 200 &  $a = b = c =   3.8598$ &   Mg(3$c$)&  0.00000 &  0.50000 & 0.50000 \\
              &              &     &        &   Sc(1$a$) &0.00000 &  0.00000 & 0.00000 \\
              &              &     &  $\alpha = \beta = \gamma =  90^{\circ}$      &  H(3$d$)  &0.00000 &  0.00000 &  0.50000  \\
 \hline
   \specialrule{0em}{1pt}{1pt}
Mg$_{3}$ScH$_{4}$    & $Pm\bar{3}m$ & 200 &   $a = b = c =   3.44577$  &  Mg(3$d$) &  0.00000  & 0.00000 &0.50000 \\
              &              &      &   &       Sc(1$b$) &  0.50000  &  0.50000 &   0.50000 \\
              &              &      & $\alpha = \beta =  \gamma = 90^{\circ}$  &  H(3$c$) & 0.00000 &   0.50000 &   0.50000 \\
              &              &      &  & H(1$a$) & 0.00000 &   0.00000  &  0.00000\\
\hline
   \specialrule{0em}{1pt}{1pt}
Mg$_{2}$ScH$_{10}$    & $C2/m$ & 200 &  $a = 4.68347$ &  Mg(4$i$) & 0.14842  &  0.50000&    0.66872 \\
              && &  $b = 2.66303$  &Sc(2$a$) & 0.00000 &   0.00000  &  0.00000\\
              &&&$c = 6.68240$&  H(4$i$) &0.01613  &  0.00000  &  0.25389 \\
              &&& $\alpha =\gamma =  90^{\circ}$ & H(4$i$)&0.08363 &   0.00000 &   0.50129 \\
              &&& $\beta =  91.8666^{\circ}$ &H(4$i$) &0.16487  &  0.50000  &  0.90258 \\
              &&&  &H(4$i$) &0.17516 &   0.50000 &   0.15337\\
              &&&  &H(4$i$) & 0.18058 &   0.50000 &   0.41760\\
\hline
   \specialrule{0em}{1pt}{1pt}
MgScH$_{2}$ & $I4_{1}/amd$ & 200  & $a = b = 3.44323$ &  Mg(4$b$)  & 0.00000 &   0.00000 &   0.50000 \\
  &&& $c = 7.13282$& Sc(4$a$) & 0.00000  &  0.00000 &   0.00000  \\
 &&& $\alpha = \beta = \gamma = 90^{\circ}$ &  H(8$e$)  & 0.00000  &  0.00000  &  0.25034  \\
\hline
   \specialrule{0em}{1pt}{1pt}
MgScH$_{3}$& $P6_{3}/mmc$ & 200 &$a = b = 2.68846$  &    Mg(2$c$)  &0.33333 &   0.66667 &   0.25000  \\
 &&&$c = 7.32296$ &Sc(2$a$) &0.00000  &  0.00000 &   0.00000\\
 &&& $\alpha = \beta = 90^{\circ}$&H(4$f$)  &0.33333  &  0.66667 &   0.88193 \\
 &&&$\gamma = 120^{\circ}$ &H(2$b$)  &0.00000 &   0.00000  &  0.25000 \\

\hline
   \specialrule{0em}{1pt}{1pt}
MgScH$_{4}$ & $P\bar{6}m2$ & 200 &$a = b = 2.81444$ &Mg(1$a$)  & 0.00000  &  0.00000 &   0.00000\\
 &&&$c = 3.55190$ &Sc(1$d$) & 0.33333  &  0.66667 &   0.50000\\
 &&&$\alpha = \beta = 90^{\circ}$ & H(2$i$)  & 0.66667  &  0.33333  &  0.25516  \\
 &&&$\gamma = 120^{\circ}$ &H(1$b$)  & 0.00000  &  0.00000 &   0.50000  \\
&&&&H(1$c$)  & 0.33333  &  0.66667  &  0.00000  \\
\hline
   \specialrule{0em}{1pt}{1pt}
MgScH$_{6}$ & $R\bar{3}m$ & 200 & $a = b =  2.66735$ &Mg(3$b$)    &-0.00000 &  -0.00000 &   0.50000 \\
 &&&$c = 13.40164$ &Sc(3$a$)   &0.00000  &  0.00000 &   0.00000 \\
 &&&$\alpha = \beta = 90^{\circ}$ &H(6$c$)    &0.00000   & 0.00000  &  0.12672 \\
 &&&$\gamma = 120^{\circ}$  &H(6$c$)    &0.00000  &  0.00000 &   0.25502  \\
&&& &H(6$c$)    &0.00000  &  0.00000  &  0.38301 \\
\hline
   \specialrule{0em}{1pt}{1pt}
MgSc$_{2}$H$_{3}$ & $Immm$ & 200 & $a = 2.54852$ &Mg(2$b$)  & 0.00000 &    0.50000 &   0.50000    \\
 &&&$b = 3.45765$ & Sc(4$j$) & 0.00000  &  0.50000  &  0.16137  \\
 &&&$c = 7.37865$& H(4$i$)  & 0.00000  &  0.00000  &  0.16481 \\
 &&&$\alpha = \beta =  \gamma = 90^{\circ}$ &H(2$c$)  & 0.00000  &  0.00000  &  0.50000 \\
\hline
   \specialrule{0em}{1pt}{1pt}
MgSc$_{2}$H$_{4}$ & $R\bar{3}m$ & 300 & $a = b = 2.57905 $  &Mg(6$c$)  &0.00000  &  0.00000  &  0.19177\\
 &&&$c = 35.63329$ &Sc(6$c$) &0.00000  &  0.00000  &  0.08687\\
 &&&$\alpha = \beta = 90^{\circ}$ &Sc(6$c$)  &0.00000  &  0.00000 &   0.36179 \\
 &&&$\gamma = 120^{\circ}$ &H(6$c$)  &0.00000  &  0.00000  &  0.13761  \\
&&&&H(6$c$)  &0.00000 &   0.00000 &   0.27601 \\
&&&&H(6$c$)  &0.00000 &   0.00000  &  0.44780 \\
&&&&H(3$a$)  &-0.00000 &  -0.00000 &   0.50000 \\
&&&&H(3$a$)  &0.00000 &   0.00000  &  0.00000 \\
\hline
   \specialrule{0em}{1pt}{1pt}
MgSc$_{2}$H$_{6}$ & $P6_{3}/mmc$ & 100 & $a = b = 2.92493$  &Mg(2$d$) &0.33333  &  0.66667  &  0.75000\\
 &&&$c = 12.42285$ &Sc(4$f$)&0.33333  &  0.66667  &  0.40589\\
 &&&$\alpha =  \beta = 90^{\circ}$ & H(4$e$) &0.00000  &  0.00000  &  0.16535\\
 &&&$\gamma = 120^{\circ}$  & H(4$f$) &0.33333  &  0.66667 &   0.55450\\
 &&& &H(2$a$) &0.00000  &  0.00000  &  0.00000\\
&&&&H(2$c$) &0.33333  &  0.66667 &   0.25000\\
\hline
   \specialrule{0em}{1pt}{1pt}
MgSc$_{2}$H$_{9}$ & $Immm$ & 200 &$a = 2.69052$ &Mg(2$d$)  &0.00000  &  0.50000  &  0.00000     \\
 &&&$b = 3.86680 $  &Sc(4$j$) &0.00000  &  0.50000  &  0.33646       \\
 &&&$c = 8.14298$ & H(8$l$)  &0.00000  &  0.26386  &  0.16221       \\
 &&&$\alpha = \beta = \gamma = 90^{\circ}$ & H(4$i$)  &0.00000 &   0.00000  &  0.32780      \\
&&&&H(4$h$)  & 0.00000  &  0.23179  &  0.50000      \\
&&&&H(2$a$)  &0.00000   & 0.00000  &  0.00000     \\
\hline
   \specialrule{0em}{1pt}{1pt}
MgSc$_{3}$H$_{12}$ & $Pm\bar{3}m$ & 200 & $a = b = c =  3.85331 $ &Mg(1$b$) & 0.50000  &  0.50000  &  0.50000    \\
 &&&$\alpha = \beta = \gamma = 90^{\circ}$ &Sc(3$d$) & 0.00000  &  0.00000  &  0.50000  \\
&&&&H(8$g$) & 0.25894  &  0.25894  &  0.25894  \\
&&&&H(3$c$) & 0.00000  &  0.50000  &  0.50000 \\
&&&&H(1$a$) & 0.00000   & 0.00000   & 0.00000  \\
\hline
\end{longtable}
\end{center}

\clearpage
\bibliographystyle{apsrev4-1}
\bibliography{references}